\documentclass[copyright,creativecommons,noderivs,noncommercial]{eptcs}

\usepackage{color}
\usepackage{amsmath}
\usepackage{amssymb}
\usepackage{graphicx}

\newtheorem{defn}{Definition}
\newtheorem{exmp}[defn]{Example}
\newtheorem{lem}[defn]{Lemma}
\newtheorem{thm}[defn]{Theorem}
\newtheorem{cor}[defn]{Corollary}

\newtheorem{prop}[defn]{Proposition}

\newenvironment{myproof}{\noindent\textit{Proof.~}}{\hfill $\boxtimes$\medskip}

\newcommand{\Par}{\mathop{||}}

\newcommand{\id}[1]{\mathit{#1}}
\newcommand{\textsosrule}[2]
{\frac{#1}{#2}}

\newcommand{\sosrule}[2]{\frac{\raisebox{.7ex}{\normalsize{$#1$}}}
                        {\raisebox{-1.0ex}{\normalsize{$#2$}}}}
\newcommand{\trans}[1]{\,{\stackrel{{#1}}{\longrightarrow}}\,}

\newcommand{\ntrans}[1]{\,{\stackrel{{#1}}{\nrightarrow}}\,}

\newcommand{\bisim}{\mathbin{\mbox{$\underline{\leftrightarrow}$}}}
\newcommand{\gbisim}[1]{\mathbin{\mbox{$\underline{\leftrightarrow}_{#1}$}}}
\newcommand{\gnbisim}[2]{\mathbin{\mbox{$\underline{\leftrightarrow}^{#2}_{#1}$}}}
\newcommand{\fhbisim}{\gbisim{\mathrm{fh}}}
\newcommand{\sbbisim}{\gbisim{\mathrm{fh}}}
\newcommand{\pfhbisim}{\gbisim{\mathrm{pfh}}}

\newcommand{\hpbisim}{\gbisim{\mathrm{hp}}}
\newcommand{\phpbisim}{\gbisim{\mathrm{php}}}
\newcommand{\hpnbisim}[1]{\gnbisim{\mathrm{hp}}{#1}}

\newcommand{\tr}[1]{\mathrel{Tr^{#1}}}
\newcommand{\bis}[1]{\mathrel{Bis^{#1}}}

\newcommand{\cibisim}{\gbisim{\mathrm{ci}}}

\newcommand{\Nat}{\,\mathbb{N}\,}

\newcommand{\Terms}[1]{\mathbb{T}(#1)}
\newcommand{\CTerms}[1]{\mathbb{C}(#1)}
\newcommand{\Eqs}[1]{\mathcal{E}(#1)}

\newcommand{\gsosprem}{\mathrm{\Delta}}

\newcommand{\vars}[1]{\mathit{vars}(#1)}

\newcommand{\ar}[1]{\mathit{ar}(#1)}

\title{Robustness of Equations Under Operational Extensions}

\author{Peter D. Mosses
\institute{Department of Computer Science, Swansea University\\
Singleton Park, Swansea, SA2~8PP, United Kingdom}
\email{p.d.mosses@swan.ac.uk}
\and
MohammadReza Mousavi\qquad Michel A. Reniers
\institute{Department of Computer Science, Eindhoven University of Technology\\
P.O.~Box~513, NL-5600~MB~~Eindhoven, The Netherlands}
\email{m.r.mousavi@tue.nl\qquad m.a.reniers@tue.nl}}

\def\authorrunning{P.D. Mosses, M.R. Mousavi \& M.A. Reniers}

\def\titlerunning{Robustness of Equations}

\begin{document}
\maketitle

\begin{abstract}
Sound behavioral equations on open terms may become unsound after
conservative extensions of the underlying operational semantics.
Providing criteria under which such equations are preserved is
extremely useful; in particular, it can avoid the need to repeat
proofs when extending the specified language.

This paper investigates preservation of sound equations for
several notions of bisimilarity on open terms: closed-instance
(ci-)bisimilarity and formal-hypothesis (fh-)bisimilarity, both
due to Robert de Simone, and hypothesis-preserving (hp-)bisimilarity,
due to Arend Rensink. For both fh-bisimilarity and hp-bisimilarity,
we prove that arbitrary sound equations on open terms are preserved
by all disjoint extensions which do not add labels. We also define
slight variations of fh- and hp-bisimilarity such that all sound
equations are preserved by arbitrary disjoint extensions. Finally,
we give two sets of syntactic criteria (on equations, resp.\
operational extensions) and prove each of them to be sufficient
for preserving ci-bisimilarity.
\end{abstract}

\section{Introduction}

Equations, pertaining to behavioral
equivalences on open terms, are not robust even under conservative extension of operational semantics
specifications, i.e., sound equations may become unsound after an operationally conservative extension (see \cite{Mousavi05-ICALP} and also examples throughout the rest of this paper).
There are several examples of this phenomenon in the literature,
for example in the domain of timed extensions of process algebras \cite{Baeten02,Reniers02} the equation $x+\delta = x$ ceases to be sound in strong bisimilarity.
Providing criteria under which
equations are preserved is extremely useful.
For example, it allows for developing general algebraic
rules for certain sub-languages -- or even for individual
constructs -- which are guaranteed to hold under all operationally conservative extensions.
This paper provides such criteria for the preservation of equations that are sound with respect to strong bisimilarity.

Note that strong bisimilarity is naturally lifted to open terms by defining two open terms to be bisimilar when each pair of their closed instantiations are bisimilar;
this is called ci-bisimilarity (for closed-instance bisimilarity \cite{deSimone85}).
In this paper, we first
recall two further notions of bisimilarity on open terms,
due to de Simone \cite{deSimone85} and Rensink \cite{Rensink00}, which are strictly finer (more distinguishing) than
ci-bisimilarity.
Subsequently, we show that a very general class of sound equalities, with respect to each of the two notions, are preserved under arbitrary disjoint extensions.
Hence, these two notions can be used to prove sound and robust
equations with respect to strong bisimilarity.
Then, we illustrate why in general ci-bisimilarity cannot be preserved under arbitrary disjoint extension,
and propose (stricter) syntactic criteria by which a certain class of axioms, or a certain class of extensions do preserve ci-bisimilarity (on open terms).

\paragraph{Related work.} In \cite{deSimone85}, de Simone studies a bisimulation proof technique for open terms
and proposes a notion of bisimulation, which is essentially the same as what we call fh-bisimilarity (for Formal Hypothesis bisimilarity) in the remainder of this paper. Rensink in \cite{Rensink00} extends the study of de Simone and provides a comparison of fh-bisimilarity
with ci-bisimilarity.
He  also proposes another notion of bisimilarity, called hp-bisimilarity (for Hypothesis Preserving bisimilarity)
and compares it to fh- and ci-bisimilarity.
In \cite{Aceto94}, Aceto, Bloom and Vaandrager give an algorithm
for generating sound and complete axioms for SOS specifications in the GSOS format of \cite{Bloom95};
they also show that the generated axioms also remain sound under certain \emph{disjoint} extensions introduced by their own algorithm.
Our results in this paper generalize and give some more insight on the  aforementioned result of \cite{Aceto94}.
In \cite{Aceto09} Aceto, Cimini and Ing\'olfsd\'ottir introduce a bisimulation proof technique for open terms called rule-matching bisimilarity, which is not generally robust under disjoint extensions;
we compare the notions studied here with rule-matching bisimilarity in an extended version of the present paper \cite{Mosses2010-TUe}.

\paragraph{Structure of the Paper.} In Section \ref{sec::pre}, we review some preliminaries from the literature.
In Section \ref{sec::fhPreserved}, we show that under some mild conditions fh- and hp-bisimilarity are preserved by conservative extensions. In Section \ref{sec::ciPreserved}, we show that the same result does not carry over trivially to ci-bisimilarity; however, we give sufficient conditions on the equations and the extensions that guarantee ci-bisimilarity to be robust. In Section \ref{sec::conc}, we conclude the paper and present some ideas for future research.

\section{Preliminaries}
\label{sec::pre}

\subsection{SOS Specifications: Syntax and Semantics}

\begin{defn}[\label{def::terms}Signatures, Terms and Substitutions]
We assume a countable set $X$ of variables. A signature $\Sigma$ is a set of function symbols (also called operators) with fixed arities;
the arity of $f$ is denoted by $\ar{f}$.
The set of terms on signature $\Sigma$, denoted by $\Terms{\Sigma}$ and
ranged over by $s, t, s_0, t_0, \ldots$, is defined inductively as follows:
variables and function symbols of arity zero (also called constants) are terms;
given a list of terms, their composition using a function symbol (while respecting the arity of the function symbol) is a term.
Terms are also called open terms;
the set of variables in $t$ is denoted by $\vars{t}$.
Closed terms on signature $\Sigma$, denoted by $\CTerms{\Sigma}$ and ranged over by $p, q, \ldots$, are those terms in $\Terms{\Sigma}$ that do not contain any variable.
A (closing) substitution $\sigma: X \rightarrow \Terms{\Sigma}$ is a function from variables to (closed) terms.
Substitutions are lifted to terms (as their domain) in the usual manner.
\end{defn}

\begin{defn}[\label{def::tss}Transition System Specification (TSS)]
A \emph{transition system specification} $T$ is a tuple $(\Sigma, L, D)$ where $\Sigma$ is a signature,
$L$ is a set of labels (with typical members $a, b, a_0, \ldots$) and
$D$ is a set of deduction rules. For all $l \in L$, and
$t, t' \in \Terms{\Sigma}$ we define that $t \trans{l} t'$ is a formula;
$t$ is its \emph{source} and $t'$ is its \emph{target}.
A formula is \emph{closed} when all terms appearing in it are closed.
A deduction rule $\id{dr} \in D$ is defined as a
pair $(H,c)$, where $H$ is a set of formulae and $c$ is a formula;
$c$ is called the \emph{conclusion} and the formulae from $H$ are called the \emph{premises}.
A deduction rule is $f$-\emph{defining} when the head operator of the source of its conclusion is $f$.
A deduction rule is an \emph{axiom} when its set of premises is the empty set.
\end{defn}

We sometimes refer to a TSS for its set of deduction rules. A deduction rule $(H, c)$ is also written as $\frac{H}{c}$;
in the latter syntax,  if the set $H$ of premises is empty, it is just left out.

\pagebreak

\begin{defn}[\label{def::ruleoid}Provable Ruloid]
A deduction rule $\frac{H}{\phi}$ is a \emph{provable ruloid} of TSS $T$
when there is a well-founded upwardly branching tree with nodes labelled by formulae and of which
\begin{itemize}
\item the root is labelled by $\phi$;
\item if a node is labelled by $\psi$ and the nodes immediately above it form the set $K$ then:
\begin{itemize}
\item $\psi$ is of the form $x \trans{a} x'$ for some distinct $x, x'\in X$,
 $\psi \in H$ and $K = \emptyset$, or
\item $\frac{K}{\psi}$ is an instance of a deduction rule in $T$.
\end{itemize}
\end{itemize}
\end{defn}

A TSS is supposed to define a transition system, i.e., a set of closed formulae.
In our setting the transition relation associated with a TSS is the set of all closed formulae $\phi$
such that $\textsosrule{}{\phi}$
is a provable ruloid.



\begin{exmp}\label{ex::basicAxioms}
As an illustration, consider a TSS $(\Sigma,L,D)$ corresponding to a sublanguage of CCS \cite{Milner89a}, where
$\Sigma$ comprises the constant 0,
a unary operator $\alpha.\_$ for each $\alpha \in L$, and
a binary operator $\_+\_$~,
$L$~is some set of actions $\{a,b,\ldots\}$, and
$D$ consists of the following deduction rules for each $\alpha \in L$.
\[
\sosrule{}{\alpha.x \trans{\alpha} x} \qquad
\sosrule{x \trans{\alpha} x'}{x + y \trans{\alpha} x'} \qquad
\sosrule{y \trans{\alpha} y'}{x + y \trans{\alpha} y'}
\]
The associated transition relation includes formulae such as $0+a.0 \trans{a} 0$, but no formula of the form $0 \trans{\alpha} p$.
The equations $x+(y+z)=(x+y)+z$, $x+y=y+x$, $x+x=x$ and $x+0=x$ are all sound (regardless of whether the rules of the first form above are included or not). The TSS can be extended to full CCS: this involves adding not only new operators and their defining rules, but also new labels (all co-actions $\bar{a},\bar{b},\ldots$ and the silent action $\tau$).
The associativity and commutativity equations for $+$ remain sound under any such extension. However, the last two equations cease to be sound, unless the (obvious) rules defining $+$ for the new labels are added too.
\end{exmp}
In Section~\ref{sec::fhPreserved} we establish theorems which guarantee preservation of sound equations by extension, under some mild conditions.

\subsection{Rule Formats}

It is customary in the meta-theory of SOS  to restrict the syntax of TSSs in order to obtain semantic results.
Such classes of TSSs with restricted syntax are called rule formats \cite{Aceto01,Mousavi07-TCS}.
One important rule format, studied extensively in the literature is
GSOS, which is due to Bloom, Istrail and Meyer \cite{Bloom95}.
Next, we define a subset of GSOS restricted to positive formulae.
We leave the generalization of our results
to the full GSOS format (which allows negative formulae as premises)
for the future.

\begin{defn}[\label{def::gsos}Positive GSOS Rule Format]
A deduction rule is in the \emph{positive GSOS format} when it is of the following form.
\[
\sosrule{\{x_i \trans{a_{ij}}  y_{ij} \mid i \in I, j \in J_i \}}{f(x_1,\ldots,x_n) \trans{a} t}
\]
where $n = ar(f)$, the variables $x_1,\ldots,x_n$ and $y_{ij}$ are all pairwise distinct, $I$ is a subset of $\{i
\mid 1 \leq i \leq n\}$, $I$ and $J_i$, for each $i \in I$, are finite index sets, and
$\vars{t} \subseteq \{x_1,\ldots,x_n\} \cup \{y_{ij} \mid i \in I, j \in J_i\}$.
A TSS is in the positive GSOS format when all its deduction rules are.

We denote by $\gsosprem$ the set
of all premises of the form $x \trans{l} x'$ for distinct $x, x'\in X$: $\gsosprem = \{ x \trans{l} x' \mid \allowbreak x,x' \in X \land x \neq x' \land l \in L\}$.
\end{defn}



\subsection{Extending SOS Specifications}

\begin{defn}[\label{def::extension}\label{def::compatibility}Disjoint Extension]
Consider two TSSs $T_0=(\Sigma_0,$ $L_0,$ $D_0)$ and $T_1=(\Sigma_1, L_1, D_1)$ of which the signatures agree on the arity of the shared function symbols.
The \emph{extension} of $T_0$ with $T_1$, denoted by $T_0 \cup T_1$, is defined as $(\Sigma_0 \cup \Sigma_1, L_0 \cup L_1,$ $D_0 \cup D_1)$.

$T_0 \cup T_1$ is a \emph{disjoint} extension of $T_0$ when each deduction rule in $T_1$ is $f$-defining for some $f \in \Sigma_1 \setminus \Sigma_0$.
\end{defn}
If both $T_0$ and $T_0 \cup T_1$ are in the
positive GSOS format, we speak of a \emph{disjoint positive GSOS extension}. Any disjoint positive
GSOS extension is also conservative, meaning that any transition that can be derived in the extended TSS for a closed term of the non-extended TSS is already derivable in the non-extended TSS \cite{Groote92}.


\subsection{Behavioral Equivalences}

A notion of behavioral congruence $\sim$ is defined w.r.t.\ the transition system associated with a TSS.
We write $T \models s \sim t$ to denote that two open terms $s, t \in \Terms{\Sigma}$ are related by $\sim$ w.r.t.\ $T$.
Next, we introduce the common notion of strong bisimilarity on closed terms as a notion of behavioral equivalence, and then present
three
extensions of it to open terms.

\begin{defn}[\label{def::bisim}Strong Bisimilarity on Closed Terms]
Given a TSS
$(\Sigma, L, D)$, a symmetric relation $R \subseteq \CTerms{\Sigma} \times \CTerms{\Sigma}$ is a {\em
strong bisimulation} when for each $(p, q) \in R$, $l \in L$ and $p'\in \CTerms{\Sigma}$, if  $p \trans{l} p'$, then
there exists a $q' \in \CTerms{\Sigma}$ such that $q \trans{l} q'$ and $(p', q') \in R$.

Two closed terms $p, q \in \Terms{\Sigma}$ are \emph{strongly bisimilar}, or just
 \emph{bisimilar}, when there exists a strong bisimulation relation $R$ such
 that $(p,q) \in R$. We write $p \bisim q$ when $p$ and $q$ are
 bisimilar, and refer to the relation $\bisim$ as \emph{bisimilarity}.
\end{defn}

It is well-known that sound
equations 
with respect to strong bisimilarity on closed terms remain sound under disjoint extensions \cite{Fokkink98}; in order to study the same result for open terms, we first need a notion of behavioral equivalence for open terms.
The following definition presents a natural extension of strong bisimilarity to open terms.
It is often just called strong bisimilarity (on open terms) in the literature,
but here, we  call it \emph{closed-instance} bisimilarity (ci-bisimilarity)
following \cite{Rensink00}, to distinguish it from the finer notions of bisimilarity presented afterwards.

\begin{defn}[\label{def::obisim}Closed-Instance Bisimilarity]
Two open terms $s, t \in \Terms{\Sigma}$ are \emph{closed-instance bisimilar}, denoted by $s \cibisim t$, when for all
closing substitutions $\sigma: X \rightarrow \CTerms{\Sigma}$, $\sigma(s) \bisim \sigma(t)$.
\end{defn}

De Simone \cite{deSimone85} introduced an alternative notion of strong bisimilarity on open terms, called \emph{formal hypothesis bisimilarity} (fh-bisimilarity). He defined it for rules in the de Simone format; the corresponding definition for rules in the positive GSOS format is as follows.

\begin{defn}[\label{def::fhbisim}FH-Bisimilarity]
A symmetric relation $R \subseteq \Terms{\Sigma} \times \Terms{\Sigma}$ is an \emph{fh-bisimulation} when for each two open terms $s, t \in \Terms{\Sigma}$ such that $(s, t) \in R$, for each provable ruloid $\frac{\Gamma}{s \trans{l} s'}$, there exists a provable ruloid $\frac{\Gamma}{t \trans{l} t'}$ such that $(s', t') \in R$.

Open terms $s$ and $t$ are {\em fh-bisimilar}, denoted by $s \fhbisim t$,
when there exists an fh-bisimulation $R$ such that $(s, t) \in R$.
\end{defn}

\begin{defn}[\label{def::sbbisim}SB-Bisimilarity]
A symmetric relation $R \subseteq \Terms{\Sigma} \times \Terms{\Sigma}$ is an \emph{fh-bisimulation}
when for each two open terms $s, t \in \Terms{\Sigma}$ such that $(s, t) \in R$,
the following two items holds:

\begin{enumerate}
\item for each $\sigma, \sigma' : X \rightarrow \Terms{\Sigma}$ such that for all $x \in X$, $(\sigma(x), \sigma'(x)) \in R$,
it holds that
$(\sigma(s), \sigma'(t)) \in R$, and
\item for each provable ruloid $\frac{}{s \trans{l} s'}$, there exists a provable ruloid $\frac{}{t \trans{l} t'}$
such that $(s', t') \in R$.
\end{enumerate}

Open terms $s$ and $t$ are {\em sb-bisimilar}, denoted by $s \sbbisim t$,
when there exists an sb-bisimulation $R$ such that $(s, t) \in R$.
\end{defn}





\begin{exmp}
Consider the TSS with the following deduction rules
\[
\sosrule{x \trans{a} x'}{x+y \trans{a} x'}
\qquad
\sosrule{y \trans{a} y'}{x+y \trans{a} y'}
\]
The open terms $x+(y+z)$ and $(x+y)+z$ are fh-bisimilar. The relation $R = \{ (x+(y+z),(x+y)+z), \allowbreak ((x+y)+z,x+(y+z)) \mid x,y,z \in X \} \cup \{ (x,x) \mid x \in X \}$ is an fh-bisimulation.
\end{exmp}

Rensink \cite{Rensink00} defined fh-bisimilarity for conditional transition systems. He also introduced a coarser (i.e., more
identifying) 
notion called \emph{hypothesis-preserving bisimilarity} (hp-bisimilarity), based on indexed families of binary relations (similar to \emph{history-preserving bisimilarity} \cite{vanGlabbeek89}). The corresponding definition for
positive
GSOS is as follows.

\begin{defn}[\label{def::hpbisim}HP-Bisimilarity]
A class of symmetric relations $\left(R_\Gamma \right)_{\Gamma \subseteq \gsosprem}$, with $R_\Gamma \subseteq \Terms{\Sigma} \times \Terms{\Sigma}$ for each $\Gamma \subseteq \gsosprem$, is an \emph{hp-bisimulation} when for each two open terms $s, t \in \Terms{\Sigma}$ and each $\Gamma \subseteq \gsosprem$ such that $(s, t) \in R_\Gamma$, for each provable ruloid $\frac{\Gamma'}{s \trans{l} s'}$ with $\Gamma \subseteq \Gamma'$, there exists a provable ruloid $\frac{\Gamma'}{t \trans{l} t'}$ such that $(s', t') \in R_{\Gamma'}$.

Open terms $s$ and $t$ are \emph{hp-bisimilar}, denoted by $s \hpbisim t$,
when there exists a hp-bisimulation $(R_\Gamma)_{\Gamma \subseteq \gsosprem}$ such that $(s, t) \in R_\emptyset$.
\end{defn}
Note that \cite{Rensink00} also defined a notion of hp-bisimilarity under a given set of hypotheses, which we will not address any further in this paper.

FH-bisimilarity implies hp-bisimilarity, which in turn implies ci-bisimilarity \cite[Theorem 3.7]{Rensink00}. The reverse implications do not hold \cite[Example 3.3]{Rensink00}.

CI-bisimilarity is not preserved by disjoint positive GSOS extensions (see e.g., \cite[Example 4]{Mousavi05-ICALP} and also Examples \ref{ex::satisfiable} and \ref{ex::satisfiable2} in the remainder of this paper).
In the next section we show that under some mild conditions the notions of fh- and hp-bisimilarity \emph{are} preserved by disjoint positive GSOS extensions.



Note that $\bisim$ and $\cibisim$ coincide on closed terms. Furthermore for TSSs in the positive GSOS format $\hpbisim$ and $\fhbisim$ on closed terms
also coincide with $\bisim$ (and hence with $\cibisim$ as well).
%

\subsection{Equational Theories}

\begin{defn}[Equational Theory]
The set of all equations over terms of signature $\Sigma$ is denoted by $\Eqs{\Sigma}$.
An \emph{equational theory} $E$ over $\Sigma$ is a subset of $\Eqs{\Sigma}$.
An equational theory $E$ is \emph{proper} if for each $t = t' \in E$, neither $t$ nor $t'$ is a variable.

An equational theory $E$ \emph{proves} an equation $t = t'$,
denoted by $E \vdash t = t'$ when $t =t'$ is in the smallest equivalence and congruence closure of $E$.

An equational theory $E$ is {\em sound} w.r.t.\ to a TSS $T$ (also on signature $\Sigma$) and
a particular notion of behavioral congruence $\sim$ if and only if
for all $t, t' \in \Terms{\Sigma}$, if $E \vdash t = t'$, then it holds that $T \models t \sim t'$.

Consider a TSS $T_0$; its (disjoint) extension $T_0 \cup T_1$
\emph{preserves} an equivalence $\sim$ w.r.t. $T_0$, when all sound equational theories
w.r.t. $\sim$ on $T_0$ are also sound w.r.t.\ $\sim$ on $T_0 \cup T_1$.
\end{defn}

\section{Disjoint Extensions Preserve FH- and HP-Bisimilarity}
\label{sec::fhPreserved}

In this section we show that both fh-bisimilarity and hp-bisimilarity are not necessarily preserved by disjoint extensions, not even for proper equations. Then we show that fh-bisimilarity and hp-bisimilarity are preserved by any disjoint extension that does not add new labels to the original TSS.
We also introduce subsets of fh-bisimilarity and hp-bisimilarity, called proper fh-bisimilarity and proper hp-bisimilarity, for which we show that they are preserved by arbitrary disjoint extensions.


\begin{exmp}
\label{ex::x-fx}
Consider a TSS $T=(\Sigma,L,D)$ with $\Sigma$ comprising a unary function symbol $f$, $L = \{a\}$ and $D$ comprising only the following deduction rule.
\[
\sosrule{x \trans{a} x'}{f(x) \trans{a} x'}
\]
Obviously, $f(x) \fhbisim x$ and therefore, since $\fhbisim \subseteq \hpbisim$, also $f(x) \hpbisim x$. Now, consider the extension with TSS $T'=(\Sigma',L',D')$ with $\Sigma'$ comprising only the constant $b$, $L' = \{b\}$ and $D'$ comprising only the following  deduction rule.
\[
\sosrule{}{b \trans{b} b}
\]
Now, it no longer holds that $f(x)$ and $x$ are hp-bisimilar and therefore they are also not fh-bisimilar. The reason is that the extension of the label set with label $b$ results in provable ruloids $\frac{x \trans{b} y}{x \trans{b} y}$, for each $x$ and~$y$. These can not be mimicked by any provable ruloids of $f(x)$.
\end{exmp}

The problem with the above example is that the extension introduces provable ruloids for terms over the old syntax, namely the variables. In fact, any equation of the form $x = f(t_1,\cdots,t_n)$ can be violated by a disjoint extension that introduces a new label (even without introducing new syntax).

The following example shows that for both fh- and hp-bisimilarity it does not suffice either to restrict the preservation result to only those equalities that are proper.

\begin{exmp}
\label{ex::x+y-y+x}
Consider a TSS $T=(\Sigma,L,D)$ with $\Sigma$ comprising a binary function symbol $+$, $L = \{a\}$ and $D$ comprising only the following deduction rules.
\[
\sosrule{x \trans{a} x'}{x+y \trans{a} x'+x'}
\quad
\sosrule{y \trans{a} y'}{x+y \trans{a} y'}
\]
Obviously, $x+y \fhbisim y+x$, and therefore also $x+y \hpbisim y+x$. Now, consider the extension with TSS $T'=(\Sigma',L',D')$ with $\Sigma'$ comprising only the constant $b$, $L' = \{b\}$ and $D'$ comprising only the following  deduction rule.
\[
\sosrule{}{b \trans{b} b}
\]
Now, it no longer holds that $x+y$ and $y+x$ are hp-bisimilar. The reason is that the hp-bisimilarity of $x+y$ and $y+x$ depends on hp-bisimilarity of $x+x$ and $x$. As in the previous example, this equation is not preserved by the extension.
\end{exmp}

One way to preserve fh- and hp-bisimilarity is to restrict the extensions to those that do not introduce any new labels, i.e., extensions which only add new function symbols and their defining rules.

\begin{thm}\label{th::nolabel}
FH-bisimilarity is preserved under any disjoint positive GSOS extension that does not add labels.
HP-bisimilarity is preserved under any disjoint positive GSOS extension that does not add labels.
\end{thm}

\begin{myproof}
We give the proof for the preservation of fh-bisimilarity. The proof for the preservation of hp-bisimilarity has the same structure and is therefore omitted.

Consider TSSs $T_0 = (\Sigma_0, L_0, D_0)$ and $T_0 \cup T_1 = (\Sigma_0 \cup \Sigma_1, L_0, D_0 \cup D_1)$ in the positive GSOS format,
where $T_0 \cup T_1$ is a disjoint extension of $T_0$.

We start with the following lemma.

\begin{lem}\label{lem::aux0}
Consider a provable ruloid $\frac{\Gamma}{s \trans{a} s'}$ w.r.t. $T_0 \cup T_1$;
if in the proof of the ruloid a deduction rule from $D_1$ is used,
then $s \in \Terms{\Sigma_0 \cup \Sigma_1} \setminus \Terms{\Sigma_0}$.
\end{lem}
\begin{myproof}
When $t \in \Terms{\Sigma_0 \cup \Sigma_1} \setminus \Terms{\Sigma_0}$ and $s \trans{a} s'$ is proved directly from premises including $t \trans{b} t'$ by instantiating a rule in $D_0$, then
$s \in {\Terms{\Sigma_0 \cup \Sigma_1}} \setminus \Terms{\Sigma_0}$
is ensured by the definition of the positive GSOS format. The result is then
straightforward by an induction on the depth of the proof.
\end{myproof}

Assume that $T_0 \models s \fhbisim t$; this means that there exists a fh-bisimulation relation
$R$ such that $(s,t) \in R$.
We show that $R$ is a fh-bisimulation relation w.r.t.\ $T_0 \cup T_1$ as well.
Consider arbitrary $s,t \in \Terms{\Sigma_0 \cup \Sigma_1}$ such that $(s,t) \in R$. Hence $s,t \in \Terms{\Sigma_0}$.
Assume that $\frac{\Gamma'}{s \trans{a} s'}$ is a provable ruloid w.r.t.\ $T_0 \cup T_1$.
We aim to show that there exists a provable ruloid of the form $\frac{\Gamma'}{t \trans{a} t'}$ w.r.t.\ $T_0 \cup T_1$ such that $(s', t') \in R$.

In case $\frac{\Gamma'}{s \trans{a} s'}$ is also a provable ruloid w.r.t.\ $T_0$ we are done since it then follows from the fact that $R$ is a fh-bisimulation which proves $T_0 \models s \fhbisim t$, that there exists a provable ruloid $\frac{\Gamma'}{t \trans{a} t'}$ w.r.t.\ $T_0$ such that $(s', t') \in R$, hence
$\frac{\Gamma'}{t \trans{a} t'}$ is a provable ruloid w.r.t.\ $T_0 \cup T_1$,
and we already have that $(s', t') \in R$.

So the case remains that $\frac{\Gamma'}{s \trans{a} s'}$ is not a provable ruloid of $T_0$. Then, as the disjoint extension $T_0 \cup T_1$ does not add labels w.r.t.\ $T_0$ it has to be the case that a deduction rule from $D_1$ has been used.
Since $s \in \Terms{\Sigma_0}$,
it follows from Lemma \ref{lem::aux0} that
in the proof of $\frac{\Gamma'}{s \trans{a} s'}$ only deduction rules from $D_0$ are used. Hence, this ruloid is provable w.r.t.\ $T_0$, which contradicts the assumption that it is not.
\end{myproof}

We obtain notions of bisimilarity that are preserved by \emph{arbitrary} disjoint extensions
(i.e., possibly introducing new labels) by restricting fh- and hp-bisimilarity to `proper' pairs of terms, as follows.

\begin{defn}[Proper FH- and HP-bisimilarity]
A pair $(s,t)$ of terms is \emph{proper} if both $s$ and $t$ are not just variables, or they are the same variable.

An fh-bisimulation $R$ is called \emph{proper} if all pairs in $R$ are proper.
Two terms $s$ and $t$ are \emph{proper} fh-bisimilar, notation $s \pfhbisim t$,  if there exists a proper fh-bisimulation $R$ that relates these terms.

An hp-bisimulation $(R_\Gamma)_{\Gamma \subseteq \Delta}$ is called \emph{proper} if all pairs in each $R_\Gamma$ are proper. Two terms $s$ and $t$ are \emph{proper} hp-bisimilar, notation $s \phpbisim t$,  if there exists a proper hp-bisimulation $(R_\Gamma)_{\Gamma \subseteq \Delta}$ such that $R_\emptyset$ relates these terms.
\end{defn}

Since a proper (fh- or hp-) bisimulation is also a plain (fh- or hp-) bisimulation, $s \pfhbisim t$ implies $s \fhbisim t$ and $s \phpbisim t$ implies $s \hpbisim t$. Examples~\ref{ex::x-fx} and~\ref{ex::x+y-y+x} illustrate the difference between proper and plain bisimilarity: in Example~\ref{ex::x-fx} we have $f(x) \fhbisim x$ but not $f(x) \pfhbisim x$ (since no proper bisimulation can contain the pair $(f(x),x)$); and in Example~\ref{ex::x+y-y+x} we have $x+y \hpbisim y+x$ but not $x+y \phpbisim y+x$ (since when $(R_\Gamma)_{\Gamma \subseteq \Delta}$ is an hp-bisimulation in that example, $(x+y, y+x) \in R_\emptyset$ implies $(x'+x', x') \in R_{\{x \trans{a} x'\}}$).

Next we show that proper fh-bisimilarity and proper hp-bisimilarity are preserved by any disjoint positive GSOS extension.

\begin{thm}\label{th::fhpres}
Proper fh-bisimilarity is preserved under any disjoint positive GSOS extension: if $T_0 \models s \pfhbisim t$ then $T_0 \cup T_1 \models s \pfhbisim t$.
\end{thm}
\begin{myproof}
Consider TSSs $T_0 = (\Sigma_0, L_0, D_0)$ and $T_0 \cup T_1 = (\Sigma_0 \cup \Sigma_1, L_0 \cup L_1, D_0 \cup D_1)$ in the positive GSOS format,
where $T_0 \cup T_1$ is a disjoint extension of $T_0$.
Assume that $T_0 \models s \pfhbisim t$; this means that there exists a proper fh-bisimulation relation $R$ such that $(s,t) \in R$.
We show that $R$ is a proper fh-bisimulation relation w.r.t.\ $T_0 \cup T_1$ as well.

Consider arbitrary $s,t \in \Terms{\Sigma_0 \cup \Sigma_1}$
such that $(s,t) \in R$. Hence $s,t \in \Terms{\Sigma_0}$. Since $(s,t)$ is proper we can distinguish two cases. The case that $s$ and $t$ are one and the same variable is trivial. For the other case assume that $(s,t)$ are both not just a single variable.
Assume that $\frac{\Gamma'}{s \trans{a} s'}$ is a provable ruloid w.r.t.\ $T_0 \cup T_1$.
We aim to show that there exists a provable ruloid of the form $\frac{\Gamma'}{t \trans{a} t'}$ w.r.t.\ $T_0 \cup T_1$ such that $(s', t') \in R$.

Since $s \in \Terms{\Sigma_0}$, and $s$ cannot be a variable since $R$ is a proper bisimulation,
it follows from (the contraposition of) Lemma \ref{lem::aux0} that
in the proof of $\frac{\Gamma'}{s \trans{a} s'}$ only deduction rules from $D_0$ are used. Hence, this ruloid is provable w.r.t.\ $T_0$.
It then follows from the fact that $R$ is a proper fh-bisimulation which proves $T_0 \models s \pfhbisim t$, that there exists a provable ruloid $\frac{\Gamma'}{t \trans{a} t'}$ w.r.t.\ $T_0$ such that $(s', t') \in R$, hence
$\frac{\Gamma'}{t \trans{a} t'}$ is a provable ruloid w.r.t.\ $T_0 \cup T_1$,
and we already have that $(s', t') \in R$.
\end{myproof}

\begin{thm}\label{th::hppres}
Proper hp-bisimilarity is preserved under any disjoint positive GSOS extension:
if $T_0 \models s \phpbisim t$ then $T_0 \cup T_1 \models s \phpbisim t$.
\end{thm}
\begin{myproof}
Consider TSSs $T_0 = (\Sigma_0, L_0, D_0)$ and $T_0 \cup T_1 = (\Sigma_0 \cup \Sigma_1, L_0 \cup L_1, D_0 \cup D_1)$
in the positive GSOS format,
where $T_0 \cup T_1$ is a disjoint extension of $T_0$.
Assume that $T_0 \models s \phpbisim t$; this means that there exists a proper hp-bisimulation
$(R_\Gamma)_{\Gamma \subseteq \Delta}$ w.r.t.\ $T_0$ such that $(s,t) \in R_\emptyset$.
We show that
$(R_\Gamma)_{\Gamma \subseteq \Delta}$ is a proper hp-bisimulation w.r.t.\ $T_0 \cup T_1$.

Consider $s, t \in \Terms{\Sigma_0}$ such that $(s,t) \in R_{\Gamma}$ for some $\Gamma \subseteq \Delta$. Since $(s,t)$ is proper we can distinguish two cases. The case that $s$ and $t$ are one and the same variable is trivial. For the other case assume that $(s,t)$ are both not just a single variable.
Assume that $\frac{\Gamma'}{s \trans{a} s'}$ with $\Gamma \subseteq \Gamma'$ is a provable ruloid w.r.t.\ $T_0 \cup T_1$.
We aim to show that there exists a term $t'$ such that $\frac{\Gamma'}{t \trans{a} t'}$ is a provable ruloid w.r.t.\ $T_0 \cup T_1$ and $(s', t') \in R_{\Gamma'}$.

Since $s \in \Terms{\Sigma_0}$ and $s$ is not a variable, it follows from Lemma \ref{lem::aux0} that
in the proof of $\frac{\Gamma'}{s \trans{a} s'}$ only deduction rules from $D_0$ are used. Hence, this ruloid $\frac{\Gamma'}{s \trans{a} s'}$ is provable w.r.t.\ $T_0$.
It then follows from $T_0 \models s \hpbisim t$ that there exists a term $t'$ such that $\frac{\Gamma'}{t \trans{a} t'}$ is a provable ruloid w.r.t.\ $T_0$ and $(s', t') \in R_{\Gamma'}$. Hence
$\frac{\Gamma'}{t \trans{a} t'}$ is a provable ruloid w.r.t.\ $T_0 \cup T_1$,
and we already have that $(s', t') \in R_{\Gamma'}$.
\end{myproof}

\section{Preserving CI-Bisimilarity}
\label{sec::ciPreserved}

\subsection{Disjoint extensions do not preserve CI-Bisimilarity}
It is well known that ci-bisimilarity is not preserved even for the disjoint extensions of TSSs.
Next, we give two abstract examples which illustrate this phenomenon and also hint at its two different causes.

\begin{exmp}\label{ex::satisfiable}
Consider the TSS with the signature containing the constant $0$ and the binary operator $+$, the set of labels $L = \{a, b, \ldots\}$ and the following set of deduction rules.
\[
\sosrule{x \trans{l} x'}{x + y \trans{l} x'} ~l \in L \quad \sosrule{y \trans{l} y'}{x + y \trans{l} y'} ~l \in L
\]
Since the only present constant is $0$, it does hold that $x + y \cibisim 0$.
Consider a disjoint extension of the above-given TSS with a constant $a$ which has the following deduction rule.
\[
\sosrule{}{a \trans{a} 0}
\]
Then $x + y\cibisim 0$ does not hold
anymore because, for example, $a + 0$ is not
bisimilar to $0$.
\end{exmp}

The equation
$x+ y = 0$ is not robust w.r.t.\ ci-bisimilarity
because the premises of $+$ are not satisfiable in the original TSS,
but become satisfiable, leading to some ``new'' behavior, in the extended TSS.

\begin{exmp}\label{ex::satisfiable2}
Consider the TSS with the signature containing a constant $a^\omega$ and the unary operator $f$, the set of labels $L = \{a, b, \ldots\}$ and the following set of deduction rules.
\[
\sosrule{}{a^\omega \trans{a} a^\omega} \quad \sosrule{x \trans{l} x'}{f(x) \trans{l} x'} ~l \in L
\]
For the above TSS it does hold that $f(x) \cibisim a^\omega$, but by adding
a constant $a$ with the deduction rule given in Example \ref{ex::satisfiable}, this
bismilarity ceases to hold.
\end{exmp}

The reason for this phenomenon is that the original language is not rich enough to generate all possible behavior;
hence although the premise of the deduction rule for $f$ is satisfied, the result of the transition of $f(x)$ is
confined to the behavior allowed by $a^\omega$ and thus, by extending the language $f(x)$ may show some new behavior.

We solve these issues in two ways: first, in Section \ref{subsec::robustEq}, we define some syntactic criteria on equations (and deduction rules for function symbols appearing in them), which guarantee that the equations remain sound under
\emph{any disjoint positive GSOS extension};
then, in Section \ref{subsec::robustEx}, we propose syntactic criteria on the deduction rules appearing in the disjoint extensions,
which guarantee that \emph{any sound equations} remain sound under such disjoint extensions.

\subsection{\label{subsec::robustEq}Robust Equations}

\begin{defn}[\label{def::nonEvolving}Non-evolving Indices]
For an $f$-defining deduction rule in the positive GSOS format of the following form,
\[
\sosrule{\{x_i \trans{a_{ij}}  y_{ij} \mid i \in I, j \in J_i \}}{f(x_0, \ldots, x_{n-1}) \trans{a} t}
\]
where $f$ is an $n$-ary function symbol, index $i < n$ is called \emph{non-evolving}, when $x_i \notin \vars{t}$ and
for each $j \in J_i$, $y_{ij} \notin \vars{t}$.


Index $i < n$ is non-evolving for function symbol $f$, if it is non-evolving for all $f$-defining deduction rules.

\end{defn}

A term appearing at a non-evolving index may be tested at the current state
but will have no influence in the future behavior of the term,
because neither itself nor its derivative (targets of its possible transitions)
can appear in the target of any transition of the current state.

\newcommand{\readyeq}{\mathrel{\approx_r}}
\newcommand{\readyAct}[2]{\mathit{initial}_{#2}(#1)}

\begin{defn}[\label{def::initFertile}\label{def::ready}Initial Action Equivalence and Initial Fertility]
Given a TSS $T = (\Sigma, L, D)$, the set of \emph{initial actions} of a process $p \in \CTerms{\Sigma}$ w.r.t.\ $T$, denoted by $\readyAct{p}{T}$, is defined by $\{ l \mid {l \in L} \land \exists_{p'\in \CTerms{\Sigma}} T \models p \trans{l} p'\}$.
Two closed terms $p, q \in \CTerms{\Sigma}$ are \emph{initial action equivalent} w.r.t.\ T
when
$\readyAct{p}{T} = \readyAct{q}{T}$.

TSS $T$ is \emph{initially fertile} when  for each $L' \subseteq L$, there is a process $p \in \CTerms{\Sigma}$ such that $\readyAct{p}{T} = L'$.
\end{defn}

Now we have all the necessary ingredients to establish when a ci-bisimulation is robust under arbitrary disjoint extensions.

\begin{thm}\label{th::nonevolving}
Assume that an equation $t = t'$, where $t, t' \in \Terms{\Sigma}$, is sound with respect to ci-bisimilarity for an initially fertile TSS $T = (\Sigma, D, R)$. If $t$ and $t'$ (individually) do not have repeated occurrences of any variable and
each open term in $t$ and $t'$ is the argument in a non-evolving index position of a function symbol $f$ from $\Sigma$,
then $t = t'$ remains sound with respect to ci-bisimilarity for any disjoint extension of $T$.
\end{thm}

\begin{myproof}
We start with the following lemmata, which show the role of non-evolving indices in our context.

\begin{lem}\label{lem::readyEqSubst}
Consider a TSS $T = (\Sigma, L, D)$, two closing substitutions $\sigma, \sigma': X \rightarrow \CTerms{\Sigma}$
and a set of terms $t_k \in \CTerms{\Sigma}$, for $k\in K$, such that, for each $k \in K$, $t_k$ does not contain repetition of variables,
each open term in $t_k$ is the argument of a non-evolving index of a function symbol (w.r.t.\ $T$)
and for each $x \in \bigcup_{k \in K} \vars{t_k}$, $\readyAct{\sigma'(x)}{T}$ $=$ $\readyAct{\sigma(x)}{T}$.
It holds that for each $t_k$, $T \models \sigma(t_k) \trans{l_k} p_k$ if and only if $T \models \sigma'(t_k) \trans{l_k} p_k$.
\end{lem}

\begin{myproof}
The lemma is symmetric in $\sigma$ and $\sigma'$ and
hence, proving the implication from left to right suffices.

We do this by an induction on the maximum depth of the proofs for
$T \models \sigma(t_k) \trans{l_k} p_k$, for all $k \in K$.
Each $t_k$ is of the form $f_k(s_{k0}, \ldots, s_{kn_k-1})$
($t_k$ cannot be a variable because open terms, and hence variables,
are only allowed to appear in the non-evolving indices of a function
symbol).
The last deduction rule applied to derive each transition $T \models
\sigma(t_k) \trans{l_k} p_k$  is of the following form:
\[
\sosrule{\{x_{ki} \trans{l_{kij}}  y_{kij} \mid i \in I_k, j \in J_i
\}}{f_k(x_{k0}, \ldots, x_{kn_k-1}) \trans{l_k} t'_k},
\]
for
a substitution $\sigma_k$ such that for each $j < n_k$,
$\sigma_k(x_{kj}) \equiv  \sigma(s_{kj})$, $\sigma_k(t'_k) \equiv
p_k$, and
$T \models \sigma_k(x_{ki}) \trans{l_{kij}}  \sigma_k(y_{kij})$
with a smaller proof.
Our goal is to define a set of substitutions $\sigma'_k$ such that
$\sigma'_k(x_{kj}) \equiv  \sigma'(s_{kj})$, $\sigma'_k(t'_k) \equiv
p_k$, and
$T \models \sigma'_k(x_{ki}) \trans{l_{kij}}  \sigma'_k(y_{kij})$.

Take the set of proofs of all premises of such rules, i.e.,
$T \models \sigma(s_{ki}) \trans{l_{kij}}  \sigma_k(y_{kij})$. Either $s_{ki}$ is a variable,
then we have that $\readyAct{\sigma'(s_{ki})}{T}$ $=$ $\readyAct{\sigma(s_{ki})}{T}$ and hence
$T \models \sigma(s_{ki}) \trans{l_{kij}}  p_{kij}$, for some $p_{kij}$.
Define $\sigma'_k(x_{ki}) \doteq \sigma'(s_{ki})$ and $\sigma'_k(y_{kij}) \doteq p_{kij}$.
Note that $s_{ki}$ appears in a non-evolving index of $f$ and hence $y_{kij}$ cannot appear in $t'_k$.
Otherwise, for the set of proofs
$T \models \sigma(s_{ki}) \trans{l_{kij}}  \sigma_k(y_{kij})$ such that $s_{ki}$ is not a variable, the induction hypothesis applies and hence,
we have that $T \models \sigma'(s_{ki}) \trans{l_{kij}}  \sigma_k(y_{kij})$.
Define $\sigma'_k(x_{ki}) \doteq \sigma'(s_{ki})$ and $\sigma'_k(y_{kij}) \doteq \sigma_k(y_{kij})$.
This way, we have completed the definition of $\sigma'_k$ substitutions satisfying the requirements set before.
By applying $\sigma'_k$ to the last deduction rule of the proof for
$T \models \sigma(t_k) \trans{l_k} p_k$, we obtain
$T \models \sigma'_k(f_k(x_{k0}, \ldots, x_{kn_k-1})) \trans{l_k} \sigma'(t'_k)$, or by the definition of $\sigma'_k$,
$T \models \sigma'(f_k(s_{k0}, \ldots, s_{kn_k-1})) \trans{l_k} \sigma'_k(t'_k)$, and by the property of $\sigma'_k$ and the structure of $t_k$,
$T \models \sigma'(t_k) \equiv \sigma'(f_k(s_{k0}, \ldots, s_{kn_k-1})) \trans{l_k} \sigma(t'_k) \equiv p_k$, which was to be shown.
\end{myproof}

\begin{lem}\label{lem::nonEvolving1}
Consider an initially fertile TSS $T = (\Sigma, L, D)$ and
a disjoint extension $T' = (\Sigma', L', D')$ of~$T$.
Consider a closing substitution $\sigma': X \rightarrow \CTerms{\Sigma'}$ and a set of terms $t_k \in \CTerms{\Sigma}$, for $k\in K$ such that, for each $k \in K$, $t_k$ does not contain repetition of variables and each open term in $t_k$ is the argument of a non-evolving index of a function symbol (w.r.t.\ $T$).
If there exists terms $p_k'\in\CTerms{\Sigma}$ and labels $l_k \in L$
  such that $T' \models \sigma'(t_k) \trans{l_k} p'_k$ for each $k\in K$, then $T \models \sigma(t_k) \trans{l_k} p'_k$
  for some $\sigma: X \rightarrow \CTerms{\Sigma}$
  such that $\readyAct{\sigma'(x)}{T'} \cap L$ $=$ $\readyAct{\sigma(x)}{T}$, for each
  $x \in \bigcup_{k \in K} \vars{t_k}$. 
\end{lem}

\begin{myproof}
We do this by an induction on the maximum depth of the proofs for
$T' \models \sigma'(t_k) \trans{l_k} p'_k$, for all $k \in K$.
Each $t_k$ is of the form $f_k(s_{k0}, \ldots, s_{kn_k-1})$
($t_k$ cannot be a variable because open terms, and hence variables,
are only allowed to appear in the non-evolving indices of a function
symbol).
The last deduction rule applied to derive each transition $T' \models
\sigma'(t_k) \trans{l_k} p'_k$  is of the following form:
\[
\sosrule{\{x_{ki} \trans{l_{kij}}  y_{kij} \mid i \in I_k, j \in J_i
\}}{f_k(x_{k0}, \ldots, x_{kn_k-1}) \trans{l_k} t'_k},
\]
and a substitution $\sigma'_k$ such that for each $j < n_k$,
$\sigma'_k(x_{kj}) \equiv  \sigma'(s_{kj})$, $\sigma'_k(t'_k) \equiv
p'_k$, and
$T'$ $\models$ $\sigma'_k(x_{ki})$ $\trans{l_{kij}}$ $\sigma'_k(y_{kij})$
with a smaller proof.
Our goal is to define a collection of substitutions $\sigma_k: X \rightarrow \CTerms{\Sigma}$  such
that for each $k \in K$, $T \models \sigma_k(t_k) \trans{l_k} p'_k$ and a substitution $\sigma: X \rightarrow \CTerms{\Sigma}$
such that $\sigma_k(x_{kj}) \equiv  \sigma(s_{kj})$, $\sigma_k(t'_k) \equiv
p'_k$, and for each   $x \in \bigcup_{k \in K} \vars{t_k}$,   $\readyAct{\sigma'(x)}{T'} \cap L$ $=$ $\readyAct{\sigma(x)}{T}$.
We make a case distinction based on the structure of each $s_{ki}$:

\begin{enumerate}
\item Either $s_{ki}$ is a closed term, then define $\sigma_k(x_{ki}) \doteq \sigma'_k(x_{ki})$  $\equiv \sigma'(s_{ki}) \in \CTerms{\Sigma}$ (because
we have that $t_k \in \Terms{\Sigma}$) and define $\sigma_k(y_{kij}) \doteq \sigma'_k(y_{kij})$, for each $j \in I_{i}$.
Then we have that $T \models \sigma_k(x_{ki}) \trans{l_{kij}}  \sigma_k(y_{kij})$.
Moreover, we have that $\sigma_k(y_{kij}) \equiv \sigma'_k(y_{kij}) \in \CTerms{\Sigma}$, because
the extension $T'$ is disjoint, hence conservative, and $\sigma_k(x_{ki})$, i.e., the source of the transition
$\sigma_k(x_{ki}) \trans{l_{kij}}  \sigma_k(y_{kij})$, is a closed term in $\CTerms{\Sigma}$.

\item or $s_{ki}$ is a variable, then $i$ is a non-evolving index of $f_k$ and $s_{ki}$ does not appear
anywhere else in $t_k$. Since $T$ is initially fertile,
there exist $p_{ki}, p'_{kij} \in \CTerms{\Sigma}$, for each $j \in J_i$ such that
$\readyAct{p_{ki}}{T} = \readyAct{ \sigma'_k(x_{ki}) }{T'} \cap L$
(following Definition \ref{def::initFertile} and
the fact $\readyAct{\sigma'_k(x_{ki})}{T'} \cap L \subseteq L$) and hence $T \models p_{ki} \trans{a_{kij}} p'_{kij}$.
Define $\sigma_k(x_{ki}) \doteq p_{ki}$ and $\sigma_k(y_{kij}) \doteq p'_{kij}$ and we have that
$T \models \sigma_k(x_{ki}) \trans{a_{kij}} \sigma_k(y_{kij})$.
Since $s_{ki}$ is a variable, it is justified to define $\sigma$ on $s_{ki}$;
define $\sigma(s_{ki}) \doteq p_{ki}$ and
it follows that $\readyAct{\sigma(s_{ki})}{T} = \readyAct{ \sigma'(s_{ki}) }{T'} \cap L$.


\item or $s_{ki}$ is an open term but not a variable,
then we have that $T' \models \sigma'(s_{ki}) \equiv \sigma'_k(x_{ki}) \trans{a_{kij}} \sigma'_k(y_{kij})$
for each $j \in J_i$ with a smaller proof tree than that of $\sigma(t)$.
Hence, for the set of all such $s_{ki}$ transitions, the induction hypothesis applies
and we know that there exists $\sigma''$ such that $T \models \sigma''(s_{ki}) \trans{a_{kij}} \sigma'_k(y_{kij})$
and for each $x \in \vars{s_{ki}}$,  $\readyAct{\sigma''(x)}{T'} \cap L$ $=$ $\readyAct{\sigma'(x)}{T}$.
For each variable $x$ in the domain of $\sigma''$, not defined by the previous item, define
$\sigma(x) \doteq \sigma''(x)$.
Note that if $\sigma$ has been defined by the previous item it holds that
$\readyAct{\sigma''(x)}{T'} \cap L$ $=$ $\readyAct{\sigma'(x)}{T}$ $=$ $\readyAct{\sigma(x)}{T'} \cap L$
and by Lemma \ref{lem::readyEqSubst}, we have that $T \models \sigma(s_{ki}) \trans{a_{kij}} \sigma'_k(y_{kij})$.
Define $\sigma_k(x_{ki}) \doteq \sigma(s_{ki})$ and $\sigma_k(y_{kij}) \doteq \sigma'_k(y_{kij})$
and we obtain a proof for $T \models \sigma_k(x_{ki}) \trans{a_{kij}} \sigma_k(y_{kij})$.

\end{enumerate}

 Note that, firstly, the last two items  define $\sigma$ on all variables in $\bigcup_{k \in K} \vars{t_k}$.
Secondly, it holds that $\sigma'(t'_k) \equiv p'_k \equiv \sigma(t'_k)$
because in the second and third cases where the definition of $\sigma(z)$ differs from $\sigma'(z)$,
$z$ cannot appear in $t'_k$ (because $i$ is a non-evolving index of $f_k$ and hence  $x_{ki}$ cannot appear in $t'_k$).
Thirdly, all premises of the deduction rule with $\sigma_k$ applied to them have proof:
those of which the source, i.e., $\sigma'_k(x_{ki})$,
was a closed term remain intact under $\sigma_k$,
and those with an open term as source appear at non-evolving indices and have a proof due to the induction hypothesis and satisfiability, as shown above, respectively.  Finally, $\sigma''(x)$ is a term in $\CTerms{\Sigma}$:
if $t_i$ is a closed term, then $\sigma''(x_i)$ is a closed term  in $\CTerms{\Sigma}$, because $t \in \Terms{\Sigma}$,
for all such $i$,  and each $j \in J_i$, $\sigma''(y_{ij})$ is a closed term in $\CTerms{\Sigma}$ because the extension of the TSS
is disjoint and thus conservative, if $t_i$ is an open term, then $\sigma''(x_i)$ is in $\CTerms{\Sigma}$ because it is so defined either by using the
induction hypothesis or by using the fact that the premise is satisfiable, and finally for all such $i$ and each $j \in J_i$, $\sigma''(y_{ij})$ is also
in $\CTerms{\Sigma}$  because its source is in $\CTerms{\Sigma}$ and the extension is disjoint and hence conservative.
This completes the proof of $T \models \sigma''(t) \trans{l} \sigma''(s) \equiv \sigma''(t) \trans{l} p'$ for a $\sigma'': X \rightarrow \CTerms{\Sigma}$ such that for each $x \in X$, $\readyAct{\sigma''(x)}{T'} \cap L = \readyAct{\sigma(x)}{T}$.
\end{myproof}

\begin{lem}\label{lem::nonEvolving2}
Consider $t \in \Terms{\Sigma}$; assume that
$t$ contains no repetition of variables and
each open term in $t$ appears in a non-evolving index with respect
to an initially fertile TSS $T = (\Sigma, L, D)$. Consider a disjoint extension $T' = (\Sigma', L', D')$ of $T$.
If $T \models \sigma'(t) \trans{l} p'$ for some $\sigma': X \rightarrow \CTerms{\Sigma}$ and $l \in L$, then $T' \models \sigma(t) \trans{l} p'$ for any $\sigma: X \rightarrow \CTerms{\Sigma'}$ such that $\readyAct{\sigma(x)}{T'} \cap L$ $=$ $\readyAct{\sigma'(x)}{T}$, for each $x \in X$.
\end{lem}

\begin{myproof}
We have to prove that for each $\sigma': X \rightarrow \CTerms{\Sigma}$,
if $T \models \sigma'(t) \trans{l} p'$ and there exists a $\sigma: X \rightarrow \CTerms{\Sigma'}$ such that $\readyAct{\sigma(x)}{T'} \cap L= \readyAct{\sigma'}{T}$, then $T' \models \sigma(t) \trans{l} p'$.
We do this by an induction on the depth of the proof for $T \models \sigma'(t) \trans{l} p'$.
The last deduction rule applied to derive this transition is of the following form:
\[
\sosrule{\{x_i \trans{l_{ij}}  y_{ij} \mid i \in I, j \in J_i \}}{f(x_0, \ldots, x_{n-1}) \trans{l} s},
\]
and $t \equiv f(t_0,\cdots,t_{n-1})$,
where $f$ is an $n$-ary function symbol and $t_i \in \Terms{\Sigma}$, for $i<n$, and there exists
a substitution $\sigma_0$ such that $\sigma_0(x_i) = \sigma'(t_i)$ for $i<n$, and $\sigma_0(s) \equiv p'$.
For each $i < n$, either $t_i$ is a closed term, then $\sigma'(x_i) \equiv \sigma'(t_i) \equiv \sigma(t_i) \in \CTerms{\Sigma}$, or it is an open term.
If $t_i$ is a variable,
then $i$ is a non-evolving index of $f$.
Since $T \models \sigma_0(x_i) \equiv \sigma'(t_i) \trans{l_{ij}} \sigma_0(y_{ij})$ for each $j \in J_i$,
it holds that $l_{ij} \in \readyAct{\sigma_0(x_i)}{T}$ and
because $\readyAct{\sigma'(t_i)}{T} = \readyAct{\sigma(t_i)}{T'} \cap L$,
it holds that $T' \models \sigma(t_i) \trans{l_{ij}} p'_{ij}$ for some $p'_{ij} \in \CTerms{\Sigma'}$ (note that
$l_{ij} \in \readyAct{\sigma(t_i)}{T'}$ and hence $\sigma(t_i)$ has a provable transition in $T'$ labelled $l_{ij}$).
If $t_i$ is not a variable, we have that $T \models \sigma_0(x_i) \equiv \sigma'(t_i) \trans{l_{ij}} \sigma_0(y_{ij})$ for each $j \in J_i$ with a smaller proof tree than that of $\sigma_0(t)$.
Hence, the induction hypothesis applies and we know that  $T' \models \sigma(t_i) \trans{l_{ij}} \sigma_0(y_{ij})$.

Next we define a new substitution $\sigma_1$ as given below.

\[ \sigma_1(x) =
\begin{cases}
\sigma(t_i) & \mbox{if~$x \equiv x_i$ for some $i \in I$}, \\
p'_{ij} & \mbox{if~$x \equiv y_{ij}$ for some $i \in I$ and $j \in J_i$ s.t.\ $t_i \equiv x_i$}, \\
\sigma'(x)  & \mbox{otherwise}.
\end{cases}
\]

Note that, firstly, the above substitution is well-defined: the cases are pairwise disjoint and for each case the mapped termed is defined before.
Secondly, it holds that $\sigma_0(s) \equiv p' \equiv \sigma_1(s)$ because in the first two cases where the definition of $\sigma_1(x)$ differs from $\sigma_0(x)$,
$x$ cannot appear in $s$: because $i$ is a non-evolving index in both cases, neither the sources of the transition, i.e., $x_i$ in case 1, nor
the target of the transition, i.e., $y_{ij}$ in case 2, can appear in $s$.
Thirdly, all premises of the deduction rule with $\sigma_1$ applied to them have proof: those of which the source, i.e., $\sigma_0(x_i)$, was a closed term remain intact under $\sigma_1$, and those with an open term as source appear at non-evolving indices and have a proof due to the induction hypothesis and satisfiability, as shown above, respectively.  Finally, note that $\sigma_1(x_i) \equiv \sigma'(t_i)$, for $i<n$, and hence,
$\sigma_1(f(x_0, \ldots, x_{n-1})) \equiv \sigma'(t)$.
This completes the proof of $T \models \sigma'(t) \trans{l} \sigma_1(s) \equiv \sigma'(t) \trans{l} p'$.
\end{myproof}

We now aim to show that if $t = t'$ is sound for ci-bisimilarity w.r.t.\ $T = (\Sigma, L, D)$, it is also sound for ci-bisimilarity w.r.t.\ any arbitrary disjoint extension $T' = (\Sigma', L', D')$.
Assume that $T \models t \cibisim t'$. Let $\sigma : X \rightarrow \CTerms{\Sigma'}$  be an arbitrary closing substitution. We must show that $T' \models \sigma(t) \bisim \sigma(t')$.

To show this,
assume that $T' \models \sigma(t) \trans{l} p$ for some $\sigma: X \rightarrow \CTerms{\Sigma'}$, $l \in L'$ and $p \in \CTerms{\Sigma'}$.
We show that $T' \models \sigma(t') \trans{l} p'$ for some $p'$ such that $T' \models p \bisim p'$.
It follows from Lemma \ref{lem::nonEvolving1} that $T \models \sigma'(t) \trans{l} p$, for some $\sigma': X \rightarrow \CTerms{\Sigma}$ such that $\readyAct{\sigma(x)}{T'} \cap L$ $=$ $\readyAct{\sigma'(x)}{T}$ for all $x \in X$.
Since $T \models t \cibisim t'$, it follows that $T \models \sigma'(t') \trans{l} p'$, for some $p'$ such that $T \models p \bisim p'$.
Using Lemma \ref{lem::nonEvolving2}, we have that $T' \models \sigma(t') \trans{l} p'$.
Moreover since bisimilarity on closed terms is preserved under disjoint extensions, we have that $T'\models p \bisim p'$.
This completes the proof of the theorem since we have that $T' \models \sigma(t') \trans{l} p'$ and $T'\models p \bisim p'$.
\end{myproof}

\subsection{\label{subsec::robustEx}Robust Extensions}

\begin{thm}\label{th::strongpres}
Let $\sim$ be an arbitrary equivalence that is defined in terms of transitions.
Consider a positive TSS $T_0$ and its disjoint extension $T_0 \cup T_1$.
A set of proper equations $E \subseteq \Eqs{T_0}$ is sound w.r.t.\  $T_0 \cup T_1$ and $\sim$, i.e., is robust under extension, if the set of
labels appearing in the conclusions of the deduction rules in $T_1$  is disjoint from the set of labels appearing in  the premises of the deduction rules in~$T_0$.
\end{thm}

\begin{myproof}
Take an arbitrary  $t = t' \in E$; it suffices to
show that for each $l \in L_0 \cup L_1$ and $t_0 \in \Terms{\Sigma_0 \cup \Sigma_1}$, a ruloid of the form
$\frac{H}{t \trans{l} t_0}$ is provable from $T_0$ if and only if the same ruloid is provable from $T_0 \cup T_1$.
(A~similar statement should hold for $t'$, the proof of which is identical to the one given above.)
We argue that no deduction rule in $T_1$ can contribute to the proof structure for $\frac{H}{t \trans{l} t_0}$.
First of all, the last deduction rule used in the proof can only be due to $T_0$ since the source of the conclusion of the ruloid is $t \in \Terms{\Sigma_0} \setminus X$. It also follows from the hypothesis of the theorem that if a deduction rule in the proof structure is in $T_0$, the proofs for its premises can only be due to deduction rules in $T_0$ since the labels of conclusions of the deduction rules in $T_1$ do not match the labels of premises of the deduction rules in $T_0$.
\end{myproof}

As a corollary of Theorem \ref{th::strongpres}, we have that if an extension satisfies the hypothesis of Theorem \ref{th::strongpres},
then it preserves ci-bisimilarity.

The following examples are examples of application of Theorem \ref{th::strongpres}.

\begin{exmp}
Consider the TSS $T_0$ with signature comprising a
unary function symbol $\alpha.\_$ for each $\alpha \in A_\tau$ (the set of all actions, co-actions and the invisible action $\tau$) and only the left-most deduction rule given below, for each $\alpha \in A_\tau$.
Assume that we extend $T_0$ with $T_1$ given by the other two deduction rules below, for each $\alpha \in A_\tau$, and a binary function symbol $\_ + \_$~.
\[
\sosrule{}{\alpha.x \trans{\alpha} x} \quad \mid \quad  \sosrule{x \trans{\alpha} x'}{x + y \trans{\alpha} x'} \quad \sosrule{y \trans{\alpha} y'}{x + y \trans{\alpha} y'}
\]
Given any notion of behavioral congruence $\sim$, it follows from Theorem \ref{th::strongpres} that all sound equations w.r.t.\ $T_0$ are also sound w.r.t.\ $T_0 \cup T_1$.
If $\sim$ is taken to be strong bisimilarity,
this is trivial to check manually since there is no sound equation w.r.t.\ $T_0$ apart from the trivial identities.
For weak (branching, $\eta$ and delay) bisimilarity, a sound set of equations in the original TSS is the following:
\[\alpha.\tau.x = \alpha.x \mbox{ for each } a \in A_\tau, \]
which remains sound in the extended setting.
\end{exmp}

\begin{exmp}
Consider the TSS $T_0$ defined by the following deduction rules and the signature comprising  unary function symbols $\alpha.\_$ for each $\alpha \in A_\tau$ and
$\_ \setminus H$ for each $H \subseteq A$.
\[
\sosrule{}{\alpha.x \trans{\alpha} x} \quad \sosrule{x \trans{\alpha} x'}{x \setminus H \trans{\tau} x' \setminus H} \alpha \in H
\]
Assume that we extend $T_0$ with $T_1$, which comprises the following deduction rule for each $a \in A$,
and the signature comprising a binary function symbol $\_ \Par \_$~.
\[
\sosrule{x \trans{a} x' \quad y \trans{a} y'}{x \Par y \trans{\tau} x' \Par y'}
\]
Fixing a notion of behavioral congruence $\sim$, it follows again from Theorem \ref{th::strongpres} that all sound equations w.r.t.\ $T_0$ are also sound w.r.t.\ $T_0 \cup T_1$.
If $\sim$ is taken to be strong bisimilarity, in the original systems a number of equations do hold, namely:
\[
\begin{array}{l}
 (\alpha . x) \setminus H = \tau. ( x \setminus H) ~~\mbox{for each } H \subseteq A, \alpha \in H \\
x \setminus H \setminus H' = x \setminus (H \cup H')
\end{array}
\]
But it is easy to check that all these equations are sound w.r.t.\ $T_0 \cup T_1$.
\end{exmp}

\section{Conclusions}
\label{sec::conc}
In this paper, we have defined several criteria under which different notions of strong bisimilarity on open terms are preserved by operationally conservative extensions. For the finer notions of bisimilarity on open terms, namely fh- and hp-bisimilarity, the criteria are quite mild and are applicable to most practical examples.
However, the preservation of the coarser notion of ci-bisimilarity calls for very strict criteria on either the equations or the extensions.

In \cite{Rensink00},
it is conjectured that $\cibisim$ and $\hpbisim$ coincide on open terms for
``most, if not all, of the standard process algebras''.
This conjecture is somewhat ambiguous, but we believe that the concept of non-evolving indices paves
the way to formalizing and proving it.
If such a conjecture is formulated and solved,
it allows one to use the admissive criteria defined for hp-bisimilarity to show that for
``most, if not all standard process algebras'' ci-bisimilarity is robust.

Also in \cite{Rensink00}, a notion of substitutive bisimilarity (acronym: sb-bisimilarity) is presented.
This notion is a combination of ci- and fh-bisimilarity (taking the derivable transitions of open terms from the empty set of premises
into  account) with an additional requirement of preservation of the bisimulation relation under instantiation (of variables with open terms).
It is worth noting that sb-bisimilarity is not preserved under operational extensions, as witnessed by our Examples \ref{ex::basicAxioms}, \ref{ex::x-fx} and \ref{ex::x+y-y+x}.
However, in \cite{Rensink00} it is proven that under some condition corresponding to our notion of initial fertility hp- and
sb-bisimilarity coincide. Hence, all our preservation results (Theorems \ref{th::fhpres} and \ref{th::nolabel}) for hp-bisimilarity carry over to sb-bisimilarity if both the original and the extended TSSs are initially fertile.
It remains to be further investigated whether sharper results for the preservation of sb-bisimilarity can be obtained.

Extending the definitions of fh- and hp-bisimilarity to other rule formats (e.g., full GSOS, tyft and ntyft) is non-trivial and it remains to be studied
whether the robustness results carry over to the extended settings.

\paragraph{Acknowledgments}

The anonymous referees provided useful comments on the submitted version of this paper.

\bibliographystyle{eptcs}

\end{document}